\documentstyle[11pt,newpasp,twoside]{article}
\def\rsun{{\rm  R_{\odot}}}
\def\msun{{\rm  M_{\odot}}}

\def\edcomment#1{\iffalse\marginpar{\raggedright\sl#1\/}\else\relax\fi}
\reversemarginpar

\begin{document}
\title{On the spin--up of neutron stars to millisecond pulsars in
 long--period binaries}
 \author{Hans Ritter}
\affil{Max--Planck--Institut f\"ur Astrophysik,
 Karl--Schwarzschild--Str. 1, D--85741 Garching bei M\"unchen, Germany}
\author{Andrew King}
\affil{Astronomy Group, University of Leicester, Leicester LE1 7RH,
 United Kingdom}

\begin{abstract}
We study the accretion efficiency of neutron stars in long--period
binaries ($P \ga 20^{\rm  d}$) which accrete from a giant companion. 
Using $\alpha$--disc models and taking into account the effect of
irradiation of the accretion disc by the central accretion light
source we derive explicit expressions for the duty cycle $d$ and
the accretion efficiency $\eta$ in terms of the parameters of the
binary system and the disc instability limit cycle. We show that the
absence of millisecond pulsars in wide binaries with circular orbits
and periods $P \ga 200^{\rm  d}$ can be understood as a consequence 
of the disc instability if the duration of the quiescent phase 
between two subsequent outbursts is at least a few decades.
\end{abstract}

\section{Introduction}

Ever since the detection of the first millisecond pulsar (Backer et
al. 1982) it has been clear that a ms-pulsar is a neutron star (NS)
which has been spun up by accretion in a close binary system (for a
review see Bhattacharya \& van den Heuvel 1991). Because the amount of
mass needed to spin up a NS to ms periods is rather small, i.e. $0.05
\msun \la \Delta M_{\rm  NS} \la 0.1 \msun$ (Burderi et al. 1999) and
the mass available from the donor is typically many times larger,
i.e. typically of order $1 \msun$, it appeared as if there was no
problem spinning up a NS to such short periods. Yet in recent years,
evidence has accumulated which shows that NS are inefficient
accretors. Thorsett \& Chakrabarty (1999) noted that in all cases
where the mass of a NS could be measured {\it with precision} it
turned out to be in the very narrow interval $1.3 \msun \la 
M_{\rm NS} \la 1.4 \msun$. Even if one makes the extreme assumption
that all NSs are born with exactly the same mass, this would imply
that none of these NSs could have accreted more than $\sim 0.1 \msun$,
despite the fact that some of these NSs have been spun up by accretion
and the mass available from the donor star was probably much larger
than $\sim 0.1 \msun$. In another context, King \& Ritter (1999) have
presented further evidence that NSs can be very inefficient accretors.
The main reason for this low accretion efficiency is the frequent
disparity between the mass transfer rate provided by the donor star
and the maximum rate at which the NS can accrete. The latter is
approximately the Eddington rate $\dot M_{\rm E} \approx 1.5 \times 10^{-5}
\msun {\rm yr}^{-1}$, whereas, depending on circumstances, the former
can be larger by many orders of magnitude. 

One of the obvious places in which NSs can be spun up to ms--pulsars
are wide binaries in which a NS accretes from a giant. Tauris \&
Savonije (1999) have recently devoted a detailed study to this
formation channel. One of their predictions is that, in principle,
ms--pulsars should form in binaries with final orbital periods of up
to $P \simeq 1000^{\rm d}$, even when taking into account that the
mass transfer rate in long--period systems ($P \ga 20^{\rm d}$)
exceeds the Eddington accretion rate. In addition, their calculations
predict a correlation between the final orbital period and the mass of
the ms--pulsar. Unfortunately, this cannot be tested at present. But
their first prediction can. And here observations show, contrary to
their prediction, that there are no ms--pulsars in wide binaries with
a circular orbit and a period $P \ga 200^{\rm d}$ (see e.g. table 1
in Taam, King \& Ritter 2000). In this context, Li \& Wang (1998) have
already noted that the accretion efficiency of the NS in such a binary
is far lower than has been assumed by Tauris \& Savonije (1999) if one
takes into account the fact that in all these wide binaries the
accretion disc around the NS is thermally unstable and accretion is
transient (King et al. 1997). This is because a) during an outburst
the mass flow rate through the disc is much larger than the mass
transfer rate from the donor, which, in turn, is already
super--Eddington if $P \ga 20^{\rm d}$, and b) because a NS cannot
accrete in excess of the Eddington rate. In fact, Li \& Wang (1998)
found that if the duty cycle of the disc instability limit cycle is
$d \la 10^{-2}$ then in systems with a final orbital period 
$P_{\rm f} \ga 100^{\rm d}$ the total mass accreted by the NS is 
$\Delta M_{\rm  NS} \la$ few $10^{-2}\msun$, too little to spin up a
NS to ms periods.

In this paper we elaborate on the idea of Li \& Wang (1998) and derive
explicit expressions for the duty cycle and the accretion efficiency
of the NS in terms of the parameters of the binary system and the
outburst cycle. Our calculations show that if the duration of the
quiescent phase between two subsequent outbursts exceeds a few
decades, the accretion efficiency becomes so small that in binaries
with a final orbital period $P_{\rm f} \ga 200^{\rm d}$, the NS cannot accrete
enough mass to become a ms pulsar.

\section{Basic model ingredients}

The binary systems for which we wish to decribe the accretion
processes are wide, long--period binaries with orbital periods of
typically $P \ga 20^{\rm d}$ and in which a neutron star primary of mass
$M_1$ accretes from a giant secondary, of mass $M_2$. The basic model
ingredients we use can be summarized as follows:

\subsection{Mass transfer}

We assume that mass transfer from the secondary is thermally (and
adiabatically) stable. In this case, the mass transfer rate can be
computed from a simple analytical approximation (e.g. Ritter
1999). Following King et al. (1997) and using the same model
parameters we can write for the mass transfer rate

\begin{equation}
-\dot{M}_2 = 7.3\times 10^{-10} \msun {\rm  yr}^{-1} (\zeta_e -
\zeta_R)^{-1} m_2^{1.7425} p^{0.9281},
\end{equation}
where $m_2 = M_2/\msun, p = P({\rm d})$ is the orbital period in days,
$\zeta_e$ the thermal equilibrium mass--radius exponent of the donor
star, and  $\zeta_R$ the mass--radius exponent of the donor's critical
Roche radius. For (thermal) stability we require $\zeta_e - \zeta_R >
0$ (e.g. Ritter 1988).

\subsection{Disc irradiation}

Accretion on to a compact star can result in significant external
irradiation of the accretion disc by the central accretion source
(e.g. van Paradijs 1996). Here we follow King \& Ritter (1998) and
assume that the effective temperature of the disc is kept above the
hydrogen ionization temperature $T_{\rm H} \simeq 6500K$ by irradiation from
the central accretion source out to a radius

\begin{equation}
R_{\rm h} = (B_1 \dot{M}_c)^{1/2},
\end{equation}
where $\dot{M}_c$ is the central accretion rate, and, for a neutron
star accretor, $B_1 = 3.9 \times 10^5$ cgs (King \& Ritter 1998).
Furthermore, we shall assume that the neutron star cannot accrete in
excess of the Eddington accretion rate $\dot{M}_{\rm E} \simeq 1.5 \times
10^{-8}\msun {\rm  yr}^{-1}$, i.e. that $\dot{M}_c \la
\dot{M}_{\rm E}$. This means that the maximum radius out to which the disc
can be kept hot by irradiation, i.e. at $T_{\rm eff} > T_{\rm H}$, is

\begin{equation}
R_{\rm crit} \simeq 7.1 \rsun (b_1\,\dot{m}_{\rm E})^{1/2}.
\end{equation}
Here $\dot{m}_{\rm E} = \dot{M}_{\rm E}/10^{-8} \msun {\rm  yr}^{-1}$ and $b_1 =
B_1/3.9 \times 10^5$ cgs. At the same time our assumption that
$\dot{M}_c \la \dot{M}_{\rm E}$ means that whenever the mass flow rate
through the disc is $-\dot{M}_{\rm d} > \dot{M}_{\rm E}$, the excess matter
must be ejected from the disc.

\subsection{The $\alpha$--disc model}

For describing the properties of the accretion disc around the neutron
star we use the Shakura \& Sunyaev (1973) $\alpha$--disc
model. Accordingly, if the vertical temperature stratification of the
disc is dominated by viscous heating, the disc viscosity is (e.g. 
Frank, King \& Raine 1992)

\begin{equation}
\nu_{\rm visc} = 2.7 \times 10^{15} {\rm cm}^2 {\rm s}^{-1} \alpha^{4/5} 
(-\dot{m}_{\rm d})^{3/10} m_1^{-1/4} r^{3/4},
\end{equation}
where $-\dot{M}_{\rm d}$ is the mass flow rate through the disc, $R$ the 
disc radius, $\dot{m}_{\rm d} = \dot{M}_{\rm d}/10^{-8} \msun {\rm
yr}^{-1}$, and $r = R/\rsun$. $\alpha$ is the viscosity parameter, here 
assumed to be constant with radius.

(4) holds only to the extent that the vertical temperature
stratification in the disc is dominated by viscous heating, hence its
dependence on $\dot{M}_{\rm d}$. If, on the other hand, the disc is
strongly heated by external irradiation, the vertical temperature
profile becomes nearly isothermal at the irradiation temperature
$T_{\rm irr}$. Taking this into account in the $\alpha$--viscosity
ansatz we find (see also King 1998)

\begin{equation}
\begin{array}{ll}
\nu_{\rm visc} & =  \alpha \frac{{\cal R}
T_{\rm H}}{\mu}(B_1\dot{M}_c)^{1/4} (G M_1)^{-1/2} R \\  &   
=  3.8 \times 10^{15}{\rm cm}^2 {\rm s}^{-1} \alpha \left(
\frac{\mu}{0.6}\right)^{-1}\left(\frac{T_H}{6500K}\right) 
(b_1\dot{m}_c)^{1/4} m_1^{-1/4} r. 
\end{array} 
\end{equation}
Here ${\cal R}$ is the gas constant and $\mu$ the mean molecular weight.

The maximum suface density which a disc can reach in the cool state,
i.e. before it becomes thermally unstable and must turn into the hot
state, is conventionally denoted by $\Sigma_B$. Taking $\Sigma_B$ from
computations of Ludwig, Meyer--Hofmeister \& Ritter (1995), the mass
accumulated before an outburst sets in can be written as

\begin{equation}
M_d \simeq 7 \times 10^{-10} \msun \; \alpha_c^{-4/5} m_1^{-0.37} 
r_{\rm d}^{3.10}{\cal F}. 
\end{equation}
Here $r_{\rm d} = R_{\rm d}/\rsun$, where $R_{\rm d}$ is the outer
radius of those parts of the disc which are involved in the outburst,
$\alpha_c$ the $\alpha$--parameter in the cool state, and ${\cal F}  <
1$ is the filling factor defined by this equation. ${\cal F} $ takes
into acount that a the onset of an outburst $\Sigma = \Sigma_B$ at
only one point in the disc and that $\Sigma < \Sigma_B$ elsewhere.

\subsection{Assumptions about the disc instability limit cycle}

We shall assume that the disc instability limit cycle in these wide,
long--period binary systems with a neutron star accretor can be
described as follows:

\begin{description}

\item[a)] During an outburst the disc is kept in the hot state by
irradiation out to the radius $R_{\rm h}(\dot{M}_c)$ given by (2). Because
$\dot{M}_c \la \dot{M}_{\rm E}, R_{\rm h}(\dot{M}_c) \la  R_{\rm h} (\dot{M}_{\rm E})$. If the
physical outer radius of the disc $R_{\rm disc} > R_{\rm h}(\dot{M}_c)$,
only the inner part with radius $R_{\rm d} = R_{\rm h}(\dot{M}_c)$ is hot,
whereas the outer part with $R_{\rm h}(\dot{M}_c) < R < R_{\rm disc}$
remains in the cold state (e.g. King \& Ritter 1998). Hereafter we
shall call $R_{\rm  d}$ the radius of the ``active'' disc. If, on the
other hand, $R_{\rm h}(\dot{M}_c) > R_{\rm  disc}, R_{\rm d} = R_{\rm
disc}$. If the mass flow rate through the outer parts of the active
disc is $-\dot{M}_{\rm d} (R_{\rm d}) > \dot{M}_{\rm E}$, then as
long as $-\dot{M}_{\rm d} (R_{\rm d}) > \dot{M}_{\rm E}, \dot{M}_c =
\dot{M}_{\rm E},$  and $R_{\rm h}(\dot{M}_c) = R_{\rm h} (\dot{M}_{\rm E})$ = const.

\item[b)] During an outburst the active part of the disc (out to
$R_{\rm d}$) is essentially emptied, i.e. the mass remaining in the
active part of the disc immediately after an outburst is insignificant
compared to the mass $M_{\rm d}$ given by (6) immediately before the
onset of an outburst.

\item[c)] During quiescence the mass of the active part of disc which
was accreted/\- ejected during the previous outburst is replenished at
the mass transfer rate $-\dot{M}_2$. This is tantamount to assuming
that if an inactive outer part of the disc exists, i.e if
$R_{\rm h}(\dot{M}_c) < R_{\rm disc}$, this outer part is stationary,
i.e. that $\dot{M}_{\rm d} (R) = \dot{M}_2$ for all  $R_{\rm
h}(\dot{M}_c) < R < R_{\rm  disc}$. 

\item[d)] We assume that the hot part of the disc (during an outburst)
is characterized by a viscosity parameter $\alpha_{\rm h}$, whereas in
the cool state (i.e. in the active part during quiescence and in the
inactive part) the viscosity parameter is $\alpha_c < \alpha_{\rm h}$.

\end{description}

\section{Disc accretion in long--period neutron star binaries}

Here we are mainly interested in systems which end their nuclear time
scale--driven mass transfer at very long orbital periods, i.e. $P \ga
200^{\rm d}$. With the analytical solution (Ritter 1999) one can show
that the corresponding initial periods at which stable mass transfer
must have started are typically $P_{\rm i} \ga 20^{\rm d}$, (the
precise value depending on whether mass transfer is conservative or
not, cf. Taam et al. 2000). It has been shown by King et al. (1997)
and Li \& Wang (1999) that at such long periods the mass transfer rate
provided by nuclear evolution of the donor is always too small for
stable disc accretion, i.e. the systems in question are transient,
going through a disc instability limit cycle.

In addition, the orbital period at which $R_{\rm disc} = R_{\rm crit}$ 
(with  $R_{\rm disc} > R_{\rm crit}$ for longer periods) is

\begin{equation}
P_{\rm  crit} = 12{\rm d} \left(\frac{R_{\rm disc}/R_{1,R}}{0.7}
\right)^{-3/2} m_1^{-0.675} m_2^{0.175} (b_1\,\dot{m}_{\rm E})^{3/4},  
\end{equation}
where $R_{1,R}$ is the critical Roche radius of the primary.
Thus, for practically all systems in question we also have 
$R_{\rm disc} > R_{\rm h}(\dot{M}_c)$. 

As we shall argue below, during most of an outburst the mass flow rate
through the outer parts of the active disc is highly super--Eddington
in the sense that $-\dot{M}_{\rm d}(R_{\rm d}) \gg \dot{M}_{\rm E}$. 
Therefore, at least initially, in these systems 
$R_{\rm d} = R_{\rm h}(\dot{M}_{\rm E})$ = const.

Now, if the viscosity at $R = R_{\rm d}$ does not depend on
$\dot{M}_{\rm d}$, and $R_{\rm d}$ itself is constant, both the mass
and the mass flow rate through the outer parts of the active disc
decay exponentially on the viscous time scale

\begin{equation}
\tau_{\rm visc} = \frac{R_{\rm d}^2}{3 \nu_{\rm visc}(R_{\rm d})}
\end{equation}
at the outer radius of the active disc. If initially 
$-\dot{M}_{\rm d}(R_{\rm d}, t = 0) \gg \dot{M}_{\rm E}$, we have $R_{\rm d}
= R_{\rm h}(\dot{M}_{\rm E}) = {\rm const.}$ as long as $-\dot{M}_{\rm d} >
\dot{M}_{\rm E}$ and the disc is essentially emptied during the
super--Eddington phase which lasts for a time 

\begin{equation}
t_{\rm  outb} \simeq \tau_{\rm  visc}(R_{\rm d}) \ln \left[ 
\frac{-\dot{M}_{\rm d}(R_{\rm d},t=0)}{\dot{M}_{\rm E}}\right].
\end{equation}
This is the behaviour to be expected if the active part of the disc is
dominated by irradiation, i.e. if $\nu_{\rm  visc}$ is given by
(5). If, on the other hand, viscous heating still dominates the
vertical temperature structure, $\nu_{\rm  visc}$ is given by
(4). Using (4) in (8) we see that $-\dot{M}_{\rm d} \propto M_d^{10/7}$,
i.e. that the mass flow rate and the mass in the active part of the
disc do not decay exponentially but rather as 

\begin{equation}
M_d(t) = M_d(0) \left[1 + \frac{3}{7}\frac{t}{\tau_0}\right]^{-3/7},
\end{equation}
where

\begin{equation}
\tau_0 = \frac{M_d(0)}{-\dot{M}_{\rm d}(0)}
\end{equation}
is the time scale on which the disc mass decays initially. If,
initially, $-\dot{M}_{\rm d} \gg \dot{M}_{\rm E}$, the disc is essentially
emptied during the super--Eddington phase which lasts in this case for
the time

\begin{equation}
t_{\rm outb} \simeq \frac{3}{7}(\tau_{\dot{M}_{\rm E}} -\tau_0),
\end{equation}
where

\begin{equation}
\tau_{\dot{M}_{\rm E}} = \tau_0 \left[\frac{-\dot{M}_{\rm d}(0)}{\dot{M}_{\rm E}}
\right]^{3/10}
\end{equation}
is the time scale on which the disc mass decreases once $-\dot{M}_{\rm 
d}$ reaches $\dot{M}_{\rm E}$. 

The duration of the quiescent phase is given by the time it takes to
replenish the matter which has been lost (accreted/ejected) from the
active part of the disc during the previous outburst. Because we have
assumed that the active disc is fed at the rate $-\dot M_2$,
irrespective of whether $R_{\rm disc}$ is larger or smaller than
$ R_{\rm d}$, the duration of quiescence is

\begin{equation}
t_{\rm q} \simeq \frac{{M}_d(0)}{-\dot{M}_2}.
\end{equation}
The total length of an outburst cycle is then 

\begin{equation}
t_{\rm cycle} = t_{\rm outb} + t_{\rm q}
\end{equation}
and the duty cycle of the outbursts

\begin{equation}
d = \frac{t_{\rm outb}}{t_{\rm cycle}}.
\end{equation}
To decide the question of how much the neutron star can accrete, the
quantity we need is the accretion efficiency $\eta$. Because the mass
accreted over one cycle is $\Delta M_{\rm accr} \simeq t_{\rm outb}
\dot{M}_{\rm E}$, whereas the mass transferred over the same time is
$\Delta M_{\rm transf} \simeq t_{\rm cycle} (-\dot{M}_2)$, we have 

\begin{equation}
\eta \simeq \frac{t_{\rm  outb}\cdot \dot{M}_{\rm E}}{t_{\rm
cycle}(-\dot{M}_2)} \simeq d\frac{\dot{M}_{\rm E}}{(-\dot{M}_2)}.
\end{equation}
Writing 

\begin{equation}
x = \frac{-\dot{M}_{\rm d}(0)}{\dot{M}_{\rm E}}
\end{equation}
we obtain

\begin{equation}
\eta \simeq \frac{\ln x}{x}
\end{equation}
if the mass flow rate decays exponentially, i.e. if (5) is used, and 

\begin{equation}
\eta \simeq \frac{7}{3} \frac{x^{0.3}-1}{x-x^{0.3}}
\end{equation}
if (4), i.e. (10) holds. In both cases, it is immediately seen that
$\eta$ is small if $x \gg 1$, i.e. if during the initial phases of an
outburst the mass flow rate is highly super--Eddington. 

\section{The mass flow rate through the outer disc}

In order to get $\eta$, we need to know $x$,
i.e. $-\dot{M}_{\rm d}(R_{\rm d},t=0)$. Assuming again that
$R_{\rm d}=R_{\rm h}(\dot{M}_{\rm E}) < R_{\rm  disc}$,  
$-\dot{M}_{\rm d}(R_{\rm d},t=0) > \dot{M}_{\rm E}$, and using (2), (5), (6)
and (8) we find 

\begin{equation}\begin{array}{ll} 
-\dot{M}_{\rm d}(R_{\rm d},t=0) = & 3.3 \times 10^{-6} \msun 
{\rm yr}^{-1} \alpha_{\rm h} \alpha_c^{-4/5} {\cal F}\\&
\left(\frac{\mu}{0.6}\right)^{-1}\left(\frac{T_{\rm H}}{6500K}\right)
(b_1\,\dot{m}_{\rm E})^{1.3} m_1^{-0.37}. \end{array}  
\end{equation}
We note that $\dot{M}_{\rm d}(R_{\rm d},t=0)$ as given in (21), and
therefore also $\eta$, still 
depends on the orbital period. The dependence on $P$ enters
in a subtle way via the filling factor $\cal F$ which itself must be a
function of $\dot M_2$ and hence via (1) of $P$. Unfortunately, this
dependence is not explicitly known and can only be determined from
full time--dependent calculations of the disc evolution. 

Replacing $\alpha_c^{-4/5} {\cal F}$ in (21) by means of (6), noting
that 

\begin{equation}
M_d = -\dot{M}_2 \cdot t_{\rm q} ,
\end{equation}
we get rid of the poorly known quantities $\alpha_c$ and $\cal F$ at
the price of introducing 
$t_{\rm q}$. This yields

\begin{equation}\begin{array}{ll}
\dot{M}_{\rm d}(R_{\rm d}, t=0) = & 7.6 \times 10^{-9} \msun {\rm yr}^{-1}
(\zeta_e-\zeta_R)^{-1} \left(\frac{\mu}{0.6}\right)^{-1}\left(
\frac{T_H}{6500K}\right)\\ &
(b_1\,\dot{m}_{\rm E})^{-1/4}m_1^{-1/2} m_2^{1.7425} p^{0.9281}t_{\rm
q}({\rm  yr}).
\end{array}
\end{equation}
Inserting typical values in (23), i.e $(\zeta_e-\zeta_R) \simeq 1,
\alpha_{\rm h} \simeq 0.2, \dot{m}_{\rm E}=1.5, m_1=1.4,m_2=1$ we find 
$x \simeq 1.25 t_{\rm q}({\rm yr})$ at $P=20^{\rm d}$ and $x \simeq
10.6 t_{\rm q}({\rm yr})$ at $P=200^{\rm d}$. This shows that if the
quiescent time of these discs is long, i.e. $ > $ a few decades, the
accretion efficiency becomes very small. Unfortunately, we do not have
any direct observational information about $t_{\rm q}$. However the
analogy with the properties of the outbursts in black hole soft X--ray
transients (BHSXTs) (see e.g. King \& Ritter 1998) suggests that the
quiescent times are likely to be very long, i.e. longer than several
decades. In the Discussion below we show that the absence of any
observed outbursts from the progenitors of pulsars in wide circular
binaries supports this conclusion. We note in passing that
$-\dot{M}_{\rm d}(R_{\rm d})$ must not be too large if this picture is
to be self--consistent: the assumption that $\nu_{\rm visc}$ is
dominated by external irradiation, i.e. that we may use (5), requires
that 

\begin{equation}\begin{array}{ll}
-\dot{M}_{\rm d}(R_{\rm h}(\dot{M}_{\rm E})) & < \frac{8\pi \sigma
T_H^4(B_1\dot{M}_{\rm E})^{3/2}}{3G M_1} \\&
< 1.2 \times 10^{-5}\msun {\rm  yr}^{-1}\left(\frac{T_H}{6500K}\right)^4 
(b_1\,\dot{m}_{\rm E})^{3/2} m_1^{-1}. \end{array} 
\end{equation}
If instead of (5) we use (4), i.e. a viscosity where the disc
temperature is self--consistently determined by viscous dissipation,
we obtain:

\begin{equation}
\begin{array}{ll}
-\dot{M}_{\rm d}(R_{\rm d},t=0) = & 2.0 \times 10^{-9} \msun {\rm
yr}^{-1} (\zeta_e-\zeta_R)^{-10/7}\alpha_{\rm h}^{8/7}\\ & 
(b_1\,\dot{m}_{\rm E})^{-25/28} m_1^{-5/14} m_2^{2.4893} 
p^{1.3259}t_{\rm q}({\rm  yr})^{10/7}. \end{array}
\end{equation}
Inserting the same typical values as above in (25) we obtain $x \simeq
0.7 t_{\rm q}({\rm yr})^{10/7}$ at $P=20^{\rm d}$ and 
$x \simeq 15 t_{\rm q}({\rm yr})^{10/7}$ at $P=200^{\rm d}$. From this
it is seen that if only $t_{\rm q} \ga $few {\rm yr}, disc accretion
becomes even more inefficient if the viscosity is due to viscous
dissipation rather than external irradiation. We also note that even
if (24) is violated initially, the disc's evolution will eventually be
dominated by external heating because $-\dot M_{\rm d}(R_{\rm h})$
decreases with time. 

Finally, in the case of irradiation--dominated viscosity (5) the duty
cycle becomes

\begin{equation}
\begin{array}{ll}
d & \simeq \frac{\tau_{\rm  visc}(R_{\rm h}(\dot{M}_{\rm E}))}{t_{\rm
q}} \ln x \\& 
\simeq 0.095 \alpha_{\rm h}^{-1}
\left(\frac{\mu}{0.6}\right)^{-1}\left(\frac{T_H}{6500K}\right)
(b_1\,\dot{m}_{\rm E})^{1/4} m_1^{1/2} t_{\rm q}^{-1}({\rm  yr})\ln x. 
\end{array}  
\end{equation}
Although at a first glance $d$ does not seem to depend on $P$ it 
nevertheless does: besides a weak dependence on $P$ via $\ln x$ there
is a significant one from the factor ${t_{\rm q}}^{-1} = -\dot M_2/
M_{\rm d}$. On the one hand  $\dot M_2$ depends explicitly on $P$
(cf (1)), on the other hand a more subtle dependence which we have 
mentioned already earlier enters via the filling factor $\cal F$ in 
(6). Nevertheless, if typical parameters are used in (26), i.e. 
$\alpha_{\rm h} \simeq 0.2$, $\dot m_{\rm E} = 1.5$ and $m_1 = 1.4$,
we have $d \simeq 0.12\; t_{\rm q}^{-1}({\rm  yr}) \cdot \ln x$. 
Therefore, if $t_{\rm q} \simeq$ many decades, $d \ll 1$.

\section{Discussion and Conclusions}

Li \& Wang (1998) have shown that the accretion efficiency of neutron
stars which accrete from a giant is very small and, in fact, becomes
too small for spinning up the neutron star to millisecond spin periods
in systems with a final orbital period $P_{\rm f} \ga 100^{\rm d}$ if
the duty cycle $d \la 10^{-2}$. Here we have worked out explicit
expressions for $d$ and the accretion efficiency $\eta$ in terms of
the  binary parameters and the quiescent time $t_{\rm q}$ of the
outburst cycle in the framework of an irradiated $\alpha$--disc model.
Our calculations show that both, $d$ and $\eta$ are small if 
$P > P_{\rm crit}$ and the mass flow rate at the onset of an outburst
is highly super--Eddington. Whether or not $d$ and $\eta$ are really
small enough to prevent neutron stars in systems with $P_{\rm f} \ga
200^{\rm d}$ being spun up to ms--periods depends crucially on the
duration of quiescence $t_{\rm q}$. If $t_{\rm q} \ga$ many decades,
as some observations of BHSXTs suggest, then both  $\eta$ and $d$ are
very small indeed. Unfortunately, determining $t_{\rm q}$ in the
framework of our simple model is not possible. For this fully
time--dependent calculations of the disc instability limit cycle would
be needed.

However there is another way that we can get a lower limit on $d$. We
know that pulsars in wide circular binaries must have descended from
SXTs. Yet we do not know of a single transient with a period longer
than 11.8$^{\rm d}$ (GRO J1744--28), even though X--ray satellites
would probably have detected an outburst from such a system anywhere
in the Galaxy within the last $\sim 30$~yr. This must mean that
outbursts are rather infrequent, i.e. $t_{\rm q}$ is long. We can make
this more precise by defining the following quantities:

Let $n$ be the frequency of X--ray outbursts from all systems in the
Galaxy, $N_{\rm PSR,obs}$ the currently known number of the pulsar
binaries in question, $N_{\rm PSR}$ the current number of the pulsar
binaries in question in the Galaxy, $\tau_{\rm prog}$ the lifetime of
a progenitor system in the phase of nuclear time scale mass transfer,
and $\tau_{\rm PSR}$ the lifetime of the pulsar in the pulsar binaries
in question. Here $\tau_{\rm prog}$ and $\tau_{\rm PSR}$ are suitable
averages over the relevant population. 

Now, the inventory of the binary pulsars in question is incomplete
because of flux limitation, dispersion and beaming, whereas an X--ray
outburst in one of the progenitor systems would probably be seen
wherever in the Galaxy the outburst occurs. We correct for this
incompletness by introducing a `filling factor' $f < 1$. In a 
stationary situation we then must have  

\begin{equation}
t_{\rm q} = {N_{\rm PSR} \over n}\; {\tau_{\rm prog}\over\tau_{\rm PSR}}
= {N_{\rm PSR,obs}\over n\;f}\; {\tau_{\rm prog}\over \tau_{\rm PSR}}.
\end{equation}
$\tau_{\rm PSR}$ must be of the order of the spin-down time scale 
$P/2\dot P \approx 10^8 ~{\rm yr}$, $\tau_{\rm prog}$, on the other hand,
is of the order of the nuclear time scale $t_{\infty}$ defined in
Ritter (1999, Eq. 14) which, for the systems in question, i.e. those
which end their evolution with a long orbital period 
$P_{\rm orb} \ga 200^{\rm d}$, is of the order $10^8~{\rm yr}$. With
$N_{\rm PSR,obs}=2$ (i.e. PSR B0820+02 and PSR J1803-2712), $f \approx
0.2$ and from the fact that we have not seen a single outburst in the
past 30~{\rm yr} this yields the lower limit
$t_{\rm q} > (60~{\rm yr})/f \ga 300~{\rm yr}$. We would also like to 
emphasize that we have adopted a very conservative value for $f$. The
real value is likely to be smaller still and thus the estimate for
$t_{\rm q}$ even larger because already the canonical beaming factor
for pulsars is of order $0.2$. 

Thus it is very likely that the lack of ms--pulsars in binaries with
white dwarf companions and $P_{\rm orb} \ga 200^{\rm d}$ is entirely a
consequence of dwarf nova--like disc instabilities in long--period
binaries.

Finally, we note that spinning up a NS in a long--period binary faces
yet another obstacle: during the long quiescent phases the NS acts as
a propeller (Illarionov \& Sunyaev 1975) and thereby is spun {\em down},
making the spin-up to ms spin periods even more difficult.   

\section{Acknowledgment}
H.R thanks the Leicester University Astronomy Group, where part of this work 
was done, for its hospitality and for support from its PPARC Short--Term 
Visitor grant.


\begin{references}

\reference Backer, D.C. Kulkarni, S.R., Heiles, C., Davies, M.M, Goss,
W.M. 1982, Nature, 300, 615

\reference Bhattacharya, D., van den Heuvel, E.P.J. 1991, Phys. Rep.,
203, 1

\reference Burderi, L., Possenti, A., Colpi, M., Di Salvo, T.,
D'Amico, N. 1999, \apj, 519, 285

\reference Frank, J., King, A., Raine, D. 1992, Accretion Power in
Astrophysics (Cambridge: Cambridge University Press)

\reference Illarionov, A.F., Sunyaev, R.A. 1975, \aap, 39, 185

\reference King, A.R. 1998, \mnras, 296, L45

\reference King, A.R., Frank, J., Kolb, U., Ritter, H. 1997, \apj,
484, 844

\reference King, A.R., Ritter, H. 1998, \mnras, 293, L42 

\reference King, A.R., Ritter, H. 1999, \mnras, 303, 253 

\reference Li, X.-D., Wang, Z.-R. 1998, \apj, 500, 935

\reference Ludwig, K., Meyer--Hofmeister, E., Ritter, H. 1995, \aap,
290, 473
 
\reference Ritter, H. 1988, \aap, 202, 93 

\reference Ritter, H. 1999, \mnras, 309, 360

\reference Shakura, N.I., Sunyaev, R.A. 1973, \aap, 24, 337

\reference Taam, R.E., King, A.R., Ritter, H. 2000, \apj, 541, 329

\reference Tauris, T.M., Savonije, G.J. 1999, \aap, 350, 928

\reference Thorsett, S.E., Chakrabarty, D. 1999, \apj, 512, 288

\reference van Paradijs, J. 1996, \apj, 464, L139


\end{references}
\end{document}